\documentclass[12pt,preprint]{aastex}
\slugcomment{To appear in ApJL}
\shorttitle{Hot Earth}
\shortauthors{Zhou, Aarseth, Lin, Nagasawa}
\begin{document}
\title{Origin and Ubiquity of Short-Period Earth-like Planets: \\
 Evidence for the Sequential-Accretion Theory of Planet
Formation}
%\title{}
\author{J.L. Zhou$^1$, S.J. Aarseth$^2$,
D.N.C. Lin$^{3,4}$\altaffilmark{*}, and M. Nagasawa$^5$}
\affil{$^1$Department of Astronomy, Nanjing University, Nanjing
210093, China; zhoujl@nju.edu.cn}
\affil{$^2$Institute of Astronomy,
Cambridge University, Cambridge CB3 0HA, UK; sverre@ast.cam.ac.uk}
\affil{$^3$UCO/Lick Observatory, University of California, Santa Cruz,
CA 95064, USA; lin@ucolick.org}
\affil{$^4$Center for Astrophysics, Harvard University, Cambridge MA
02138, USA}
\affil{$^5$National Astronomical Observatory Japan, Tokyo 181-8588,
Japan; nagaswmk@cc.nao.ac.jp}

\altaffiltext{*}{Corresponding should be addressed to
D.L.(lin@ucolick.org)}

\begin{abstract}
The formation of gas giant planets is assumed to be preceded by the
emergence of solid cores in the conventional sequential-accretion
paradigm. This hypothesis implies that the presence of earth-like
planets can be inferred from the detection of gas giants. A similar
prediction cannot be made with the gravitational instability
(hereafter GI) model which assumes that gas giants (hereafter giants)
formed from the collapse of gas fragments analogous to their host
stars.  We propose an observational test for the determination of the
dominant planet-formation channel. Based on the sequential-accretion
(hereafter SA) model, we identify several potential avenues which may
lead to the prolific formation of a population of close-in earth-mass
($M_\oplus$) planets (hereafter close-in earths) around stars with 1)
short-period or 2) solitary eccentric giants and 3) systems which
contain intermediate-period resonant giants. In contrast, these
close-in earths are not expected to form in systems where giants
originated rapidly through GI.  As a specific example, we suggest that
the SA processes led to the formation of the 7.5 $M_\oplus$ planet
around GJ 876 and predict that it may have an atmosphere and envelope
rich in O$_2$ and liquid water. Assessments of the ubiquity of these
planets will lead to 1) the detection of the first habitable
terrestrial planets, 2) the verification of the dominant mode of
planet formation, 3) an estimate of the fraction of
earth-harboring stars, and 4) modification of bio-marker signatures.
\end{abstract}


\section{Introduction}
The search for habitable planets is essential in the quests to unravel
the origin of the Solar System and find life elsewhere. Extra solar
planets with mass ($M_p$) down to that of Neptune \citep{marcy05,
mayor} are thought to be gas giants because some have densities
comparable to that of Jupiter. Extrapolation on the fraction of stars
which host the yet-to-be-detected $M_\oplus$ terrestrial planets
depends on the assumed models of planet formation. If giants were
formed through gas accretion onto pre-assembled cores \citep{pollack},
they would be accompanied by many earth-like planetary siblings.
However, in systems where giants were formed through GI \citep{boss},
such an association would be coincidental.

Both models were introduced and modified to account for the observed
properties of giants and extra solar planets.  Several solutions
\citep{hubickyj, alibert} have been suggested to bypass the
protoplanetary growth bottlenecks associated with the early SA models.
Likewise, photoevaporation has been adopted into the GI model
\citep{bossb} to address the giants' internal structure and
correlation with metal-rich stars.  While these revisions have led to
a more comprehensive understanding of the planet-formation process,
they have also reduced the capability to falsify the models.

We present a robust observational test to differentiate the SA and GI
models. In \S2, we briefly recapitulate their essential features.
Based on the SA model, we identify in \S3 several potential avenues
which may lead to the prolific formation of a population of close-in
earths around stars with 1) short-period, 2) solitary eccentric gas
giants, or 3) intermediate-period resonant planets.  As a specific
example, we suggest in \S4 that the dynamical structure of the system
(two giants with 30 and 60 day orbits plus a 7.5 $M_\oplus$ planet
with a 2-day orbit) around GJ 876 (with a mass $M_\ast = 0.32
M_\odot$) was established when the migrating resonant giants induced
dynamical evolution of residual embryos and led to the formation of
the close-in earth.  Since the embryos of this planet formed outside
the snow line, it is mainly made of water and it probably has an
oxygen-rich atmosphere.  In \S5, we summarize the dominant SA
processes which promote the formation of close-in earths.  Since these
planets are not expected to form if the giants originated through GI
\citep{boss}, they provide an observable test for the SA and GI
models.

\section{Differences between SA and GI models}
In the SA paradigm, planetesimals coagulate via runaway growth and
become protoplanetary embryos (with mass $M_e$) by oligarchic growth
\citep{kokuboida} on a time scale $\tau_{\rm growth}$. In gaseous
disks, the embryos' growth are stalled \citep{lissauer} when they have
swept clean the planetesimals within their feeding-zone of full width
$\Delta_{\rm fz} \sim 10-12 R_h$, where $R_h = (M_e/ 3 M_\ast )^{1/3}
a$ and $a$ are their Hill's radius and semi major axis.  For a fiducial
surface density $\Sigma_{d, g}=\Sigma_{o d, og} f_{d, g} ( {a / 1 {\rm
AU} } )^{-3/2}$ [the subscripts $d$, $g$ refer to the heavy elements
and gas with scaling factors $f_d$ and $f_g$ with respect to the
minimum mass nebular model (hereafter MMN) for the mass distribution
inside 1 and 10 AU around the Sun \citep{hayashi}], the isolation mass
is
\begin{equation}
M_{\rm iso} = 0.12 f_d^{3/2} \gamma_{\rm ice} ^{3/2}
(\frac{\Delta_{\rm fz}}{10 R_h})^{3/2} (\frac{a}{\rm AU})^{3/4}
(\frac{M_*}{M_\odot})^{-1/2} M_\oplus.
\label{eq:miso}
\end{equation}

The normalizations, $\Sigma_{od} = 10 \gamma_{\rm ice} {\rm g \
cm}^{-2}$ and $\Sigma_{og} = 2.4 \times 10^3 f_{\rm dep} {\rm g \ cm
}^{-2}$ are determined by the volatile-ice enhancement [interior
($\gamma_{\rm ice}=1$) and exterior ($\gamma_{\rm ice}=4$) to the snow
line \citep{idalin,idalin2} at $a_{\rm ice} \simeq 2.7
(M_\ast/M_\odot)^2$ AU] and the gas-depletion factors at time $\tau_\ast$
since the onset of star formation [$f_{\rm dep}={\rm exp} (-{
\tau_\ast/ \tau_{\rm dep}})$ where the gas depletion time scale
$\tau_{\rm dep}$ is observed \citep{hartmann2} to be a few
Myr].

Prior to severe gas depletion, the embryos' eccentricities ($e_e$'s)
are damped \citep{ward} on a time scale $\tau_{\rm e, damp} \simeq 300
f_g ^{-1} f_{\rm dep} ^{-1} (M_e / M_\oplus)^{-1} (a/1 {\rm AU})^2$ yr
$ < < \tau_{\rm growth}$.  Isolated embryos emerge on a time scale
$\tau_{\rm iso} = 3 \tau_{\rm growth}$, where
\begin{equation}
\tau_{\rm growth} \simeq 0.12 \gamma_{\rm ice}^{-1}
f_d^{-1} f_g^{-2/5} \left( {a \over 1 {\rm AU} } \right)^{27/10}
\left( {M_{\rm iso} \over M_\oplus} \right) ^{1/3} \left( {M_\ast \over
M_\odot} \right)^{-1/6} {\rm Myr}.
\label{eq:tgrowth}
\end{equation}

In disks with $\tau_{\rm growth} < \tau_{\rm dep}$, the embryos'
growth at $a<a_{\rm ice}$ is limited by $M_{\rm iso} < < M_\oplus$.
Just outside $a_{\rm ice}$, $f_d$ is enhanced by the grains'
interaction with the disk gas \citep{stevenson}, with $M_e > M_{\rm
core} (\sim 3-10 M_\oplus$) required for efficient gas accretion
\citep{hubickyj} so that primary giants emerge there.  However, in
low-$\Sigma_d$ disks, no giants can form because $\tau_{\rm
growth} (M_{\rm core}) > \tau_{\rm dep}$ even with a local $f_d$
enhancement.  In modest-$\Sigma_d$ disks, when $\tau_{\rm growth}
(M_{\rm core}) \sim \tau_{\rm dep}$, the rate of gas accretion and the
asymptotic $M_p$ are limited by the global disk depletion or local gap
formation \citep{bryden99}.  Beyond the gap, grains and planetesimals
accumulate which promote the emergence of secondary giants with a
delay of $\Delta \tau$ \citep{bryden00}.

During the active disk-evolution phase, giants also migrate
\citep{idalin} on a time scale
\begin{equation}
\tau_{mig} \simeq 0.8 f_g ^{-1} f_{\rm
dep} ^{-1} (M_p / M_J) (M_\odot/M_\ast) (10^{-4}/\alpha) (a/1 {\rm
AU})^{1/2} {\rm Myr},
\label{eq:taumig}
\end{equation}
where $\alpha$ is an {\it ad hoc} scaling parameter for angular
momentum transfer efficiency. In persistent disks, extensive migration
leads to short-period planets \citep{linbod}, but in rapidly
depleting disks, they may be stalled.  Multiple planets formed in
slowly evolving disks (with $\Delta \tau > \tau_{mig}$) attain wide
separations ({\it eg} Ups And). In rapidly evolving disks (with
$\Delta \tau < \tau_{\rm dep}$ and $<\tau_{\rm mig}$) the migration of
successive emerging planets lead to resonant capture \citep{pealelee}
({\it eg} GJ 876 and 55 Can).  As they continue to migrate, the
giants' eccentricities ($e_p$'s) are excited by their resonances and
damped by their tidal interaction with the gas beyond the gap
\citep{kley} on a time scale $\tau_{\rm p, damp} \sim \tau_{mig}$.

In the SA hypothesis, giants form outside the snow line while many
embryos with $M_{\rm iso}$ remain in the inner disk.  In the GI model,
giants must form in massive disks which can efficiently cool
\citep{gammie, durisen}.  These conditions are only satisfied in the
optically-thin outer regions of disks around stars with $\tau_\ast <
0.1$ Myr.  Since the dynamical time scale of such regions and the
growth time scale of the instability are $\sim 10^3$yr, which is $< <
\tau_{\rm growth}$, the GI scenario requires terrestrial planets to
form well after the giants \citep{bossb}.  In disks with $f_g > >
1$, the efficiency of angular momentum transfer is strongly enhanced
by the growth of unstable modes \citep{laugh, armitage} so that
$\alpha > 10^{-2}$ and $\tau_{\rm mig} \sim 10^3$ yr. Giants' tidal
interaction with gas in their co-orbital region of massive disks may
also lead to rapid migration \citep{masset}, though the magnitude of
this effect remain uncertain \citep{angelo}.  Migrating giants would
not capture each other into the observed mean-motion resonances
(hereafter MMR) if their $\tau_{\rm mig}$ is shorter than their
resonant libration time scale \citep{idalin9}.

\section{Emergence of close-in earths}
With a series of simulations, we show that the embryos formed prior
and interior to the giants are induced to migrate, collide, and
evolve into close-in earths.  We compute the evolution after the
formation of the giants with a fourth-order time-symmetric Hermite
scheme \citep{kokubo, aarseth} which includes the total force due to
the star, planets, and disk \citep{mardling, makiko}.  The $\Sigma_g$
distribution is assumed to be that of the MMN with a gap (of width
no less than the giants' $R_h$ and epicycle amplitude).  We
adopt values of $\tau_{\rm mig}$, $\tau_{\rm depl}$, $M_\ast$, and
$M_p$ to be within the range of the observed values \citep{hartmann2,
marcy05}.  We highlight some generic features with three
representative models:

\begin{itemize}
\item a) a Jupiter-mass ($M_J$) giant formed at $a=5.2AU$ with
$e_p=0$ (due to planet-disk tidal interaction \citep{papa}) around a
$1M_\odot$ star,

\item b) two giants [with initial $(M_p, a, e_p)= (1.67 M_J, 1.5
{\rm AU}, 0.40)$ and $(3.1 M_J$, $4.17 {\rm AU}$, $0.24)$] around a $1
M_\odot$ star,

\item c) a pair of giants [with initial $(0.56 M_J, 3.3 {\rm AU},
0)$ and $(1.89 M_J, 5.5 {\rm AU}, 0)$] around a $0.32 M_\odot$ star,
\end{itemize}

\noindent
to represent systems around HD209458, $\mu $ Ara, and GJ 876
respectively.

In Model a), we adopt $f_g=1$ and $f_d=1.25$ which yield 27 embryos
with initial $M_{\rm iso} = 0.1-0.26 M_\oplus$ and separation
$\Delta_{\rm fz} =10 R_h$ for $a=0.49-2.10$ AU.  (In all models,
$M_{\rm iso} \sim M_{\rm core}$ and $\tau_{\rm growth} (M_{\rm core})
< \tau_{\rm dep}$ at $a>a_{\rm ice}$.) The orbits of embryos exterior
to this range are rapidly destabilized by the giant whereas
$M_{\rm iso}$ of the close-in embryos does not contribute
significantly to the total mass ($M_{\rm tot} = 4.35M_\oplus$) of the
population.

\begin{figure}
\plotone{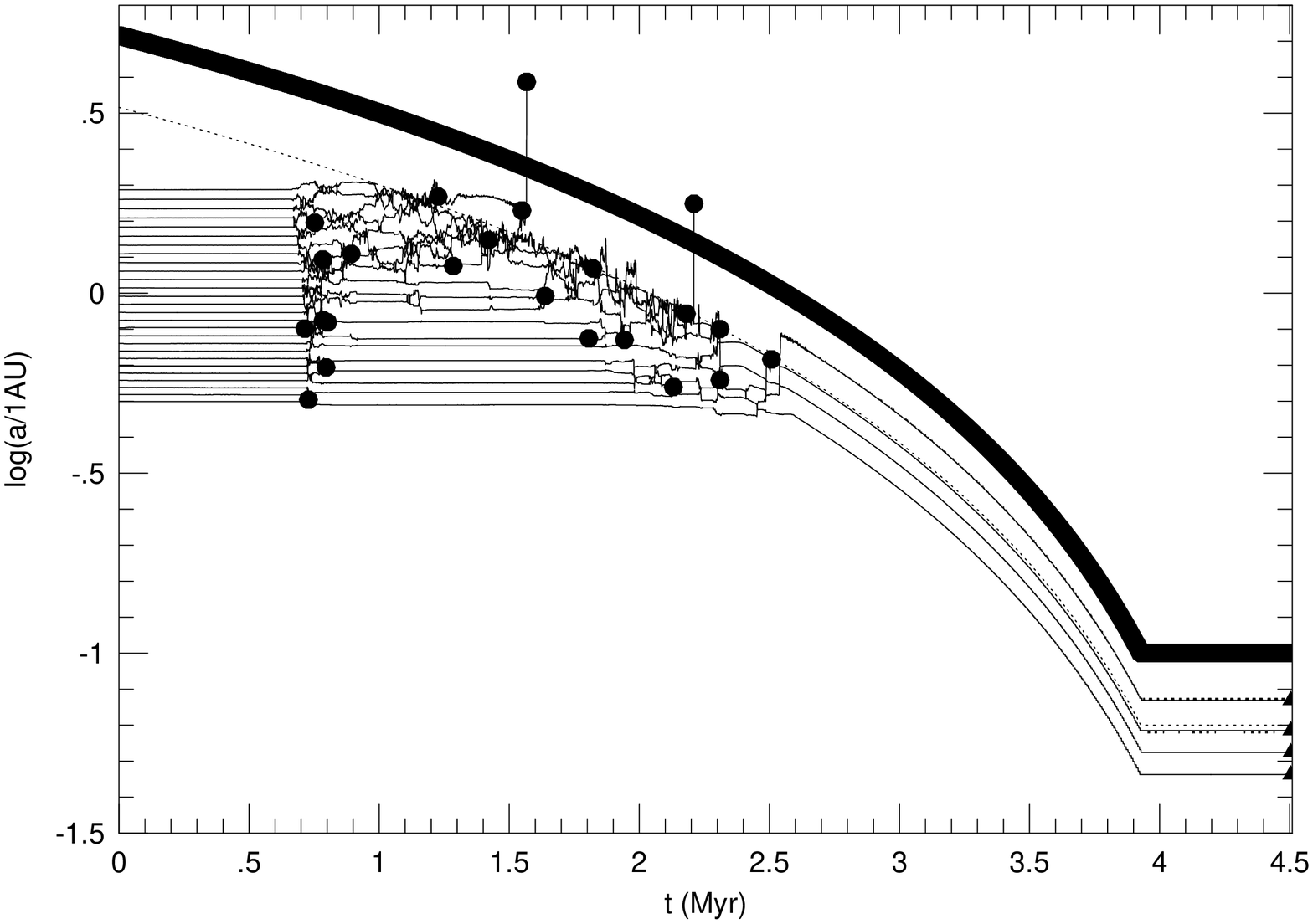} \caption{
A solitary migrating giant captures residual embryos
interior to the snow line onto its MMR.
The time evolution of the semi major axes of the giant and
 embryos for Model a) is plotted with heavy and light lines.
 The dots and triangles represent collisions and final configurations,
 and the dotted line shows the location of 2:1 resonance with the planet.
 Under the combined influence of $e_e$-damping
and the planetary resonant perturbation,
 the captured embryos migrate, collide, and grow.} \label{fig:1}
\end{figure}

The giant is assumed to migrate over $\tau_{mig} \simeq 1.8$Myr
($\alpha=10^{-4}$) and to stall at 0.1 AU.  Embryos along the giant's
migrating path are captured by its MMR's, migrate with it, and merge
into four bodies with $M_p =0.11$, $0.3$, $1.09$, \& $1.92 M_\oplus$
just inside its asymptotic 2:1 and 3:2 MMR's.  Some captures also
occur through the resonant interaction among the embryos. Since
$e_p=0$, this solitary giant's secular perturbation on the embryos
is negligible.  Due to the embryos' internal tidal dissipation
\citep{novak} and relativistic precession \citep{mardling2}, we
anticipate a slight additional $a_e$ decay and predict that
short-period giants ({\it eg} HD 209458b) are accompanied by
close-in earths (Figure \ref{fig:1}).

\begin{figure}
\plotone{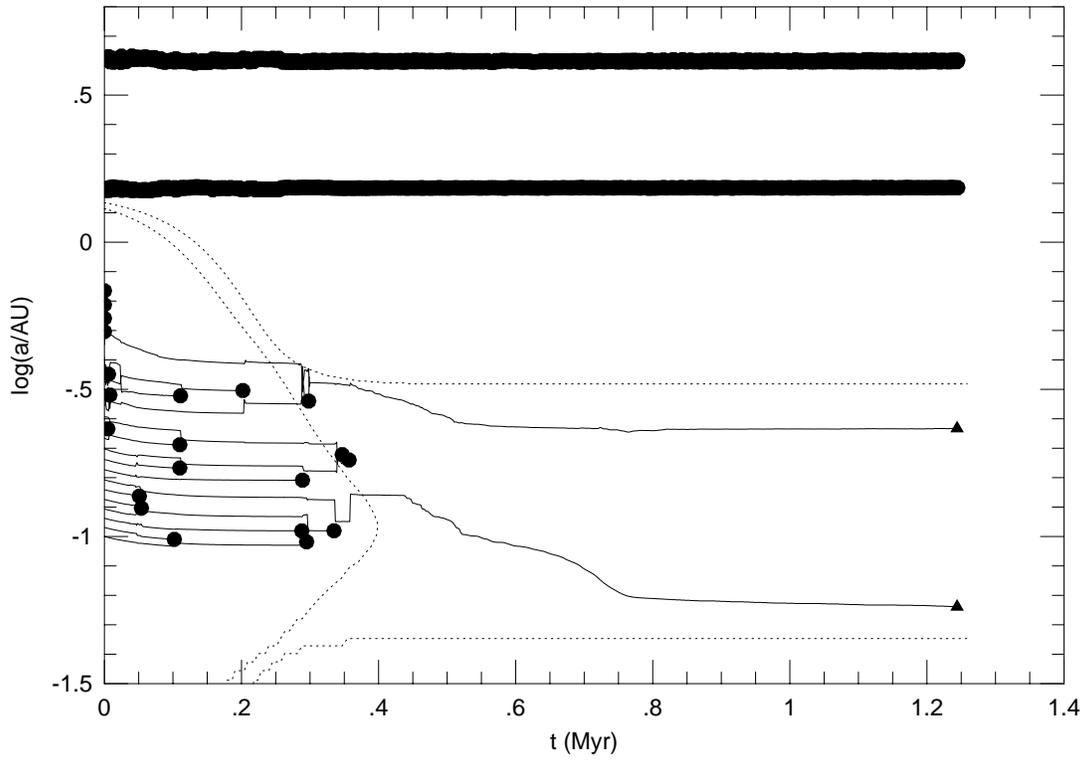} \caption{ Embryos' migration and collisions induced
by the giants' SR in Model b). As SR's (dotted lines) sweep inward
\citep{ward01} from $a_p$ and outward from the origin over
$\tau_{\rm dep}$ , the migrating embryos collide and grow in agreement
with theory.  They stall when the SR is significantly weakened or
$\dot{\varpi}_e$ is dominated by the relativistic correction. }
\label{fig:2}
\end{figure}

In Model b), we choose $f_g=f_d=4$, and $\tau_{\rm dep} (\sim 0.1
{\rm Myr} < \tau_{mig})$ with $\alpha=10^{-4}$.  The fraction of T
Tauri stars with such massive disks \citep{beckwith, hartmann} is
comparable to that of nearby stars with known planets.  The
initial 26 embryos with $M_{\rm iso} = 0.23-1.25 M_\oplus$ (a
total of $ 14.7 M_\oplus$) are separated by $\Delta_{\rm fz}=12
R_h$ at $a=0.09-1.2$ AU. The secular interaction between the two
planets (with $M_j$ and $a_j$) introduces a precession in their
longitudes of periapse $\varpi_p$ (initially orthogonal) with two
eigenfrequencies which are also modified by the disk's potential
\citep{ward, makiko}.  Due to similar effects and the star's
post-Newtonian gravity, the embryos (with $a_e$ and mean motion
$n_e$) also precess with the frequency
\begin{equation}
\dot\varpi_e =  n_e [\sum_{j=1,2}
 \frac{M_j }{4 M_\ast}  \alpha_j^2 b_{\frac{3}{2}}^{(1)}(\alpha_j)
+ 2\pi f_{\rm dep} Z(k)\frac{\Sigma_g a_e^2}{M_*} +
   3  (\frac{n_e a_e}{c})^2 ] .
\label{emt}
\end{equation}
Here $\alpha_j=a_e/a_j$, $c$ is the speed of light, and the power
index of $\Sigma_g$ in MMN is $k=3/2$, for which the constant
$Z(k)=-0.54$.  Secular resonance (hereafter SR) occurs at
locations $a_s$ where $\dot\varpi_e= \dot\varpi_p$. The resonant
embryos' $e_e$ are excited by the nearly aligned periapse
$(\varpi_e-\varpi_p$) and damped by the gas drag which also leads to
orbital decay \citep{makikolin}.  All precession frequencies and $a_s$
decrease with the disk depletion over $\tau_{\rm dep}$, leading to a
sweeping SR. As their orbits cross, the embryos coagulate into two
bodies with ($M_e,a$)$\sim (7 M_\oplus, 0.23 $AU) and ($4 M_\oplus,
0.052 $AU) (Figure \ref{fig:2}).

In Model c), we adopt $f_g =f_d=1$, $\tau_{\rm dep} =1$ Myr, and
$\tau_{mig} = 1$ Myr ($\alpha=10^{-3}$) for the outer planet.  The
initial 24 embryos (with $0.04 M_\oplus < M_{\rm iso} <1.49
M_\oplus$ and a total $\sim 7.2 M_\oplus$ separated by
$\Delta_{\rm fz} = 12 R_h$ between 0.07--0.72 AU) extend outside
the snow line ($a_{\rm ice}=0.28$ AU). With an external disk, the
outer planet undergoes orbital decay, captures the inner planet
onto its MMR, and they continue to migrate together. The resonant
excitation of the outer planet's $e_p$ is damped by the outer disk
\citep{kley} over $\tau_{\rm p damp} \sim 0.1 \tau_{mig}$. The
inner planet's damping rate is assumed to be reduced by the gas
depletion in the gap.

\begin{figure}
\plotone{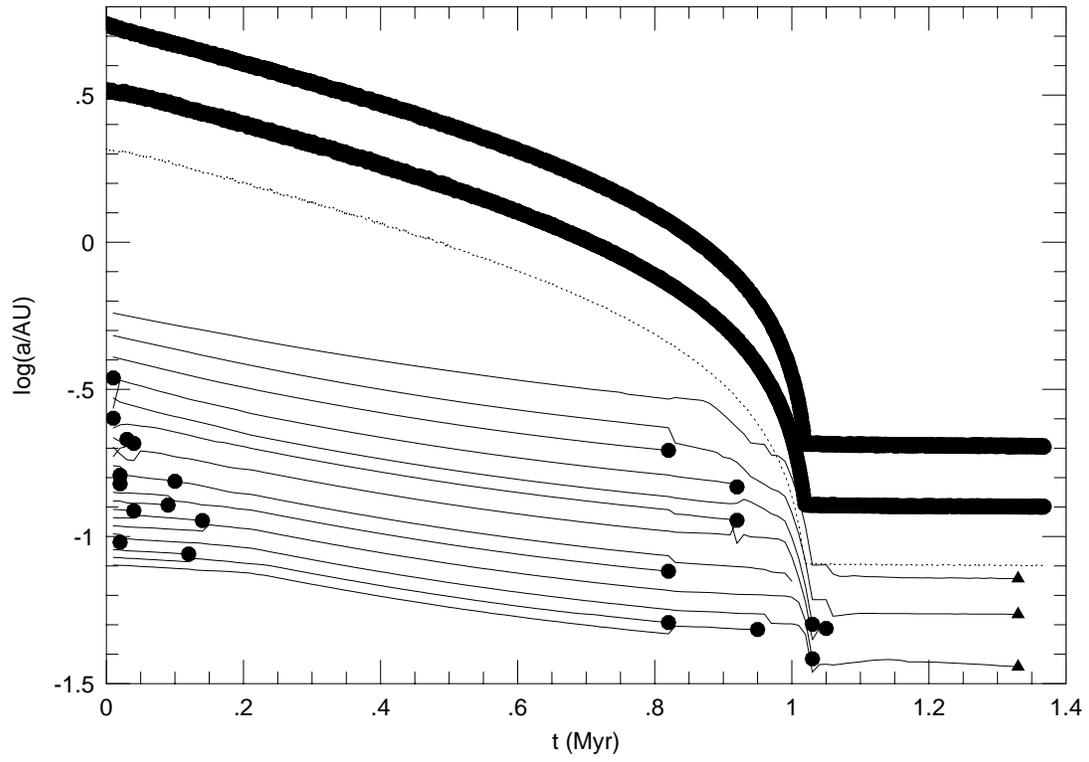} \caption{
The embryos' resonant capture and forced evolution by
migrating resonant planets in Model c).  As their orbits cross,
the embryos undergo cohesive collisions and grow.
 The dotted line shows the location of 2:1 resonance with the inner planet.} \label{fig:3}
\end{figure}

Although the embryos' initial $a_e < < a_p$, their $e_e$'s are excited
by the giants' perturbation and damped by their interaction with
the disk.  These effects lead to $a_e$ decay \citep{makikolin} on a
time scale $\sim e_e^{-2} \tau_{\rm e, damp} > \tau_{\rm mig}$. As
$a_p$ approaches their $a_e$'s, embryos are captured into the gas
giants' MMR's.  The resonant embryos are induced to 1) migrate, 2)
capture other embryos, and 3) coagulate over $\tau_{\rm mig}$.  After
the giants are stalled at 0.2 and 0.12 AU, they continue to excite
$e_e$ and reduce $a_e$ through their secular perturbation and
resonance, similar to Model b).  After 1.4 Myr, three embryos remain
with ($M_e,a$)$\sim (1.5 M_\oplus, 0.072 $AU), ($5 M_\oplus, 0.054
$AU), and ($0.7 M_\oplus, 0.036 $AU) (Figure \ref{fig:3}).  Since we
anticipate the giants' SR to sweep inward, embryos' $e_e$-damping to
weaken, migration to slow, and coalescence to occur, we continue the
calculation with a reduced $\tau_{\rm dep}$ (0.1 Myr) and find two
remaining embryos after 1.7 Myr with (6.5$M_\oplus$ , 0.056AU) and
(0.7$M_\oplus$ , 0.037AU).  The giants' secular perturbation on
the embryos is suppressed by the relativistic precession which limits
$e_e \sim 0.01$.  The embryos undergo fractional orbital decay
as they tidally interact with the host star during its life span
\citep{mardling2}.

Extensive model analysis will be presented elsewhere.  In general, the
embryos' migration driven by the giants' MMR and SR are robust
(independently of $f_g$, $\tau_{\rm dep}$, $\alpha$).  However, the
asymptotic values of $M_e$ and $a_e$ are determined by $f_d$,
$\tau_{\rm e, damp}$, $M_p$, the asymptotic values of $e_p$ and $a_p$
(which regulate $e_e$ and embryos' response).

\section{The origin of the close-in earth around GJ 876}
We choose Model c) to represent the initial disk structure around GJ
876. Both $f_g$ and $f_d$ in Model c) are high for disks around
low-mass stars \citep{hartmann}. If $M_{\rm disk} \propto M_\ast$
({\it i.e.}  $f_d \propto M_\ast$), $M_{\rm iso} < < M_\oplus$ and
$\tau_{\rm growth} > > \tau_{\rm dep}$ such that giants would rarely
form around low-mass stars \citep{idalin2}.  The total mass of
refractory material interior to $a_{\rm ice}$ is $\sim f_d M_\oplus$.
If the newly discovered planet is mostly composed of refractory
material ($\sim 1.5\%$ of all that in GJ 876), $f_d$ would need to be
$>8$ with a corresponding $M_{\rm disk}> 0.02 (f_g /f_d) (a/1 {\rm
AU})^{1/2} M_\ast$ which is gravitational unstable at $a> 1-2 AU$.

Model c) reproduces the observed properties of GJ 876.  The total
amount of solids (mostly volatile ice) within 1AU is $\sim 7-8
M_\oplus$ and the condition for giant formation ($\tau_{\rm
growth} (M_{\rm core}) < \tau_{\rm dep}$ see \S2) is satisfied at
$\sim 3$AU (where the disk is gravitationally stable).  This model
suggests that the close-in earth was formed outside the snow line,
mostly of water.  While the detectable atmospheric H$_2$O is
photo-dissociated by the stellar UV flux (Leger et al. 2004) and
hydrogen atoms escape, the atmospheric oxygen atoms are separated from
the silicate core by a deep ocean.  This process may lead to
non-biogenic formation of O$_2$, confusing the uniqueness of its
bio-marker \citep{desmarais}.

\section{Summary}
The ubiquity of close-in earths is inferred from the SA hypothesis
because their formation is promoted by: 1) the anticipated
formation of embryos prior to the emergence of giants, 2) the
driven migration of the embryos by the giants' MMR and SR, 3) the
tidal interaction between the embryos and their nascent disks, and
4) the embryos' induced collisions along their migration path.
These effects are especially important for stars with multiple
intermediate-period resonant planets.  In the GI scenario, embryos
cannot form prior to the giants.  The giants' orbits evolve too
rapidly for them to capture other giants and embryos onto their
MMR's.  Their detection through frequent radial velocity
\citep{narayan} and sub-milli magnitude transit observations of
known short-period and resonant giants could be used to
extrapolate the probability of finding terrestrial planets around
solar-type stars \citep{idalin}.

A general $\Sigma_{d,g}$ distributions (other than MMN) may lead to
diverse $M_{\rm iso}$, $\tau_{\rm growth}$, and planetary
configurations including earths with $a_e$ exterior to the $a_p$ of
giants \citep{Raymond}. Provided $\Sigma_{d,g}$ does not increase
rapidly with $a$, close-in earths are expected to be associated with
short-period giants.

In Model c), $a_{\rm ice}$'s around M stars are much closer to those
around G stars (in models a and b). We infer from the SA scenario a
prolific production of water-rich close-in earths around these
low-mass stars. Although their day-side temperature may exceed 500K,
the night-side of these close-in earths may be much cooler
\citep{burkert} because their atmosphere is likely to be covered by
opaque clouds.  Direct searches for these conspicuous close-in earths
may reveal that terrestrial planets are ubiquitous and lead to the
detection of habitable environment among them.  However, the
possibility of a physical origin for O$_2$ requires a reassessment of
the biological implications for its future detection.

\acknowledgments
We thank G. Laughlin and S. Vogt for useful conversations.  This work is
supported by NASA (NAGS5-11779, NNG04G-191G), JPL (1228184), NSF
(AST-9987417), NSFC (10233020 \& NCET-04-0468), MEXT (MEXT 16077202),
KITP \& IPAM.

\clearpage

\end{document}